# Density functional investigations on 2D-Be$_2$C as an anode for alkali Metal-ion batteries

Hetvi Jadav, Sadhana Matth, and Himanshu Pandey*

*Condensed Matter & Low-Dimensional Systems Laboratory, Department of Physics,
Sardar Vallabhbhai National Institute of Technology, Surat, Gujarat, 395007, India*

*Corresponding Author E-mail: hp@phy.svnit.ac.in*

## Abstract

Metal-ion batteries are in huge demand to cope with the increasing need for renewable energy, especially in automobiles. In this work, we apply first-principle calculations to examine two-dimensional beryllium carbide (2D-Be$_2$C) as a possible anode material for metal-ion (Na and K) batteries. 2D-Be$_2$C is a semiconductor and becomes metallic by adsorbing metal ions. Negative adsorption energy indicates stable adsorption on the monolayer of Be$_2$C. Alkali metal diffusion barrier and optimum path for minimum energy are studied within the framework of the climbing image nudged elastic band method. Here, six intermediate images are considered between the initial and final states. The lowest diffusion barriers for a single adsorbed Na and K atom are 0.016 and 0.026 eV, respectively. A maximum open circuit voltage of around 1 V is computed for K ions, whereas 0.5 V is for Na ions. Also, the maximum storage capacity of the Be$_2$C monolayer is estimated at 1785 Ah/kg.

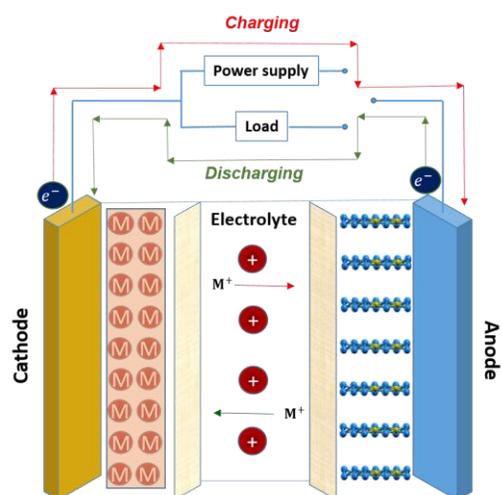

## Introduction

Conventional energy sources, such as coal and oil, have limited resources and our environment is also severely polluted by toxic gases (CO$_2$, SO$_2$, NO$_2$, *etc*.) due to the extensive usage of these old energy sources in automobiles and various industries. These gases ultimately affect human health and also cause profound climate change. Therefore, clean and sustainable energy resources like hydro, solar, wind, and rechargeable metal-ion batteries (MIBs) have been boosted to reduce the use and reliance on conventional energy sources [1-4]. Rechargeable batteries, among other energy sources, such as Ni/Cd cells, Zn/Mn cells, and lead acid batteries, are gaining popularity due to their portability, extended cycle life, quick charging time, environmental friendliness, better capacity, and higher open circuit voltage [5-7]. Due to these advantages, the most widely investigated and commercially utilized lithium-ion batteries (LIBs) are used in mobile phones, cameras, other portable electronic gadgets, electric vehicles, *etc*. [8-12].

Despite the use of LIBs in various portable and small-to-mid-scale appliances, its anticipation for large and industrial scale could not be achieved. However, LIBs offer a specific capacity of 3860 Ah/kg [13], a specific power density of around 135 W/kg [14], and a battery operating voltage of 4 V [15]; due to enormous production, a shortage of Li resources and its supply chain can be possible in the near future [16]. Safety issues, such as overheating, often damage the device [17]. Thus, to minimize the need for lithium, it is essential to use more abundant elements to investigate new batteries that are safer, more affordable, and have longer cycle lives. This is why MIBs- like sodium-, potassium-, magnesium-, and calcium-ion batteries are actively studied due to their similar electrochemical principles, higher abundance than LIBs, and, most importantly, lower prices [18-25]. Among these alkali ions, Na and K share chemical and physical characteristics with Li and belong to the same primary group. Using Na and K for MIBs will reduce the dependency on Li, which is heavily used in current LIBs. Sodium-ion batteries (NIBs) and potassium-ion batteries (KIBs) are gaining a considerable amount of interest due to economic viability and adequate natural abundance of Na and K in the crust of the Earth [25, 27]. Since oceans hold about 96.5% of all Earth's water, the supply of Na from the seawater is almost unlimited. Potassium is the 7th most abundant element in the Earth's crust; about 35 million tonnes/year is mined. The standard electrode potential of K is close to Li, whereas, for Na, this value is lower than Li [28, 29]. This feature also makes the use of Na and K worthwhile. Besides this, NIBs and KIBs can be cost-effective, as Al-foil is presumed at both electrodes of the battery for the current collection purposes [30], but the time and extended cycle life, among other factors, limit the better performance of MIBs [31]. The performance of MIBs is significantly affected by the choice of materials used for the electrodes. Previous research has shown that the characteristics and performance of an electrode material significantly affect battery working efficiency and other related parameters, such as the diffusion rate, electrode potential, energy density, and specific capacity [32-34]. Graphite is considered the most commonly used anode material for commercial LIBs; however, its applications are limited due to low capacity (372 Ah/kg) [35-37]. Even though graphite is a proven effective anode for LIBs, its use as an anode material for NIBs and KIBs is limited due to the challenges associated with metal-ion intercalation. As a result, considerable effort is required to explore the appropriate electrode materials for both KIBs and NBIs. In recent years, 2D nanomaterials have become increasingly helpful for electrochemical storage devices and have gained considerable interest for their use in MIBs due to fast redox reactions and more robust alkali metal adsorption [38-45].

In essence, rechargeable batteries comprise two electrodes with opposite polarity which are separated by a separator, into which an electrolyte solution containing dissociated salts is infused. The chemical processes occurring at the active electrodes directly affect the battery capacity and performance. The composition and characteristics of the electrode materials have a significant impact on battery performance. Therefore, exploring new electrode materials is crucial to meeting the growing demand [46, 47]. Thus far, developing high-performance electrode materials, improving the functioning of current materials, and identifying the source of electrochemical activity have all relied heavily on theoretical computation [2, 48, 49].

In the present work, we have investigated the performance of 2D-$Be_2C$ for their potential application as an anode material in MIBs using *ab-initio* calculations. We have optimized the dynamical and energy stability of the 2D-$Be_2C$ structure. Subsequently, the adsorption energy of alkali metals is computed for typical occupation sites. Electronic structure studies have examined the conductivity of $Be_2C$ monolayers. We have also computed the maximum voltage available at zero current, known as open circuit voltage and the charge storage capacity to assess the possible use of the 2D-Be₂C as an anode material for MIBs.

**Computational Details**

DFT investigations on 3×3×1 Be₂C monolayer for MIBs were carried out by utilizing the Quantum Espresso package [50] with PBE-GGA exchange-correlation functional [51]. Before relaxing the structure, various parameters, such as energy cut-off and $k$-points, were optimized via self-consistent force calculations. The plane-wave cut-off energy was set to 40 Ry to calculate the wave functions, and the charge density cut-off to 280 Ry. Non-self-consistent calculations were performed to estimate the density of states (DOSs). The electronic properties were calculated for a $k$-point grid of 15×15×1, whereas for structure relaxation and all other calculations, a 5×5×1 $k$-point grid was taken. To remove the interaction between adjacent layers, we introduced a vacuum of 20 Å. Using the DFT-D2 approach, the interaction of van der Waals was also considered. Energy barriers for the diffusion of alkali metal atoms on the surface of Be₂C were estimated by using the climbing image nudged elastic band (CI-NEB) approach via finding the energy-minimum pathway between the specified beginning and final configurations [52].

## Results and Discussion

### Structural optimization of $Be_2C$ monolayer

Figure 1(a-d) shows the 3×3×1 super-cell of 2D-Be₂C where the Be atoms bend to two distinct atomic planes located 0.46 Å above and below the C atomic plane. The top and side views of the monolayer are shown in Figs. 1(a) and (b), respectively. The dashed box in Fig. 1(a) represents the unit cell of Be₂C. Three possible adsorption sites are designated as the bottom of the beryllium (A), the top of the beryllium (B), and the top of the carbon (C). Figure 2(a) depicts the total ground state energy and its fitting using the Birch-Murnaghan equation [53], which estimates the optimized lattice parameters $a = b = 2.98$ Å, and of C-Be and Be-Be bond lengths as 1.78 Å and 1.95 Å, respectively. These values are in good agreement with the previous work [54]. The structural and thermodynamic stability of the structure is understood in terms of the formation energy $(E_f)$, which can be calculated for Be₂C as given by Eq. (1).

$$E_f = E_{Be_2C} - E_{Be} - E_C \qquad (1)$$

Where $E_{Be_2C}$, $E_{Be}$, $E_c$ are the converged ground state energies of the 2D-Be₂C, beryllium atom and carbon atom, respectively. A negative value of $E_f$ (-7.73 Ry/unit cell) suggests the thermodynamic stability of these Be₂C monolayers.

Additionally, the *ab-initio* molecular dynamics (AIMD) [55] simulations were run within the NVT ensemble to examine and verify the thermal stability of the 2D-Be₂C monolayer. The MD simulation was conducted at 300 K with a total time scale of 5 ps using the 2×2×1 supercell. Fig. 2(b) displays the top snapshot of the 2D-Be₂C monolayer structure obtained at the beginning and finish of the MD simulation, together with the total energy fluctuation as a function of time. The 2D-Be₂C monolayer's thermal stability is indicated by the 2D-Be₂C monolayer's capacity to preserve its monolayer structure despite a noticeable little deviation in monolayer.

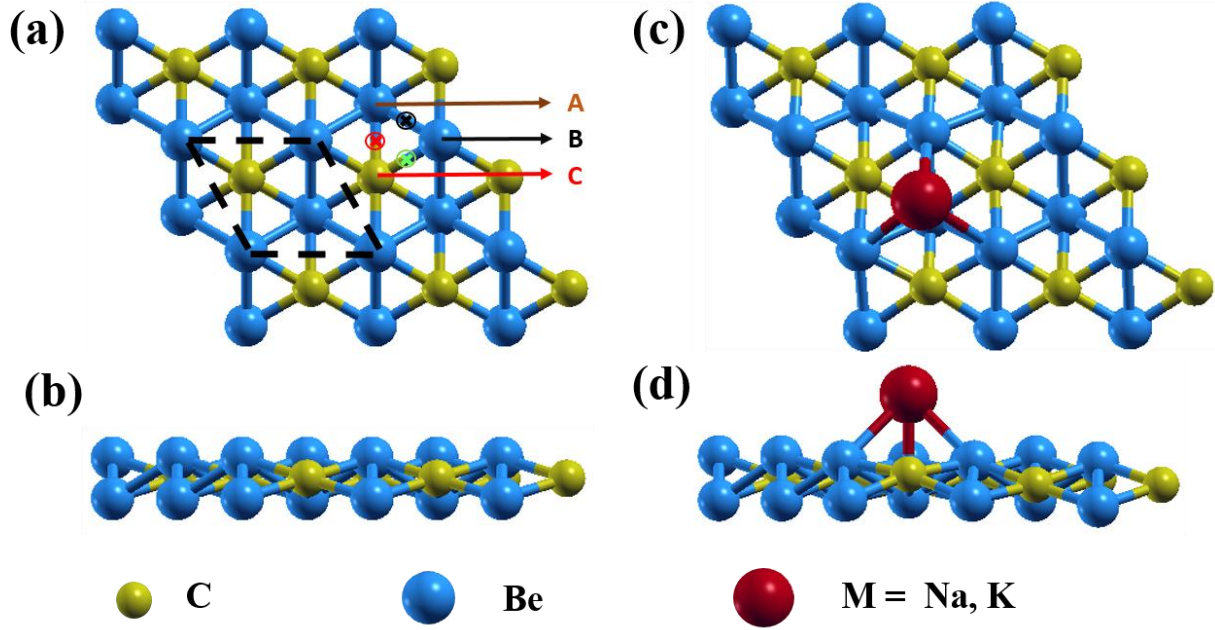

**Figure 1**: (a) Top view and (b) side view of optimized structure of 2D-Be$_2$C. In top view black dashed box indicate unit cell and capital letters (A = Bottom beryllium, B = Top beryllium, C = Carbon) show the possible adsorption sites; (c) Top and (d) side view of preferred (site-A) adsorption site for Metal atom (M = Na and K). Xcrysden program is used for structure visualization.

To understand the electron-phonon interaction and the thermodynamic stability of Be$_2$C monolayers, phonon dispersion is estimated for 2×2×1 $q$-grid, and the phonon band structure is shown in Fig. 3 (a). Since there is no evidence of negative frequencies, the phonon dispersion curves indicate that the Be$_2$C monolayers are dynamically stable. A total of nine vibrational modes for the three atoms in a primitive unit cell is observed. The top six modes with high frequencies are the optical modes. In contrast, the lower frequency mode is the out-of-plane mode, with nearly quadratic dependence at the point. The in-plane longitudinal acoustic (LA) and transverse acoustic (TA) modes have higher frequencies with linear dispersion at Γ point. It has three acoustic modes: At the high-symmetry points K and M, the degeneracy is absent, and just before and after the M point, the two crossings of the LA and TA branches can be seen.

**Electronic Properties of the Be$_2$C Monolayer**

A good conducting nature is essential for any material to work as an electrode. We study the electronic structures of the 2D-Be$_2$C after the adsorption of alkali metal atoms. Figure 3 (b-d) shows the band structure and total DOS for pristine Be$_2$C and alkali metal adsorbed Be$_2$C monolayer. Notably, the pristine monolayer of Be$_2$C has a direct semiconducting bandgap of 1.52 eV, comparable to previously reported results. [54] By adsorbing alkali metals (Na and K) atoms, a metallic characteristic appears, which is the essential requirement for an ideal battery electrode. Near the Fermi level, similar band profiles and DOS plots are observed for alkali metal adsorptions, as these atoms have the same number of electrons as the 2D-Be$_2$C.

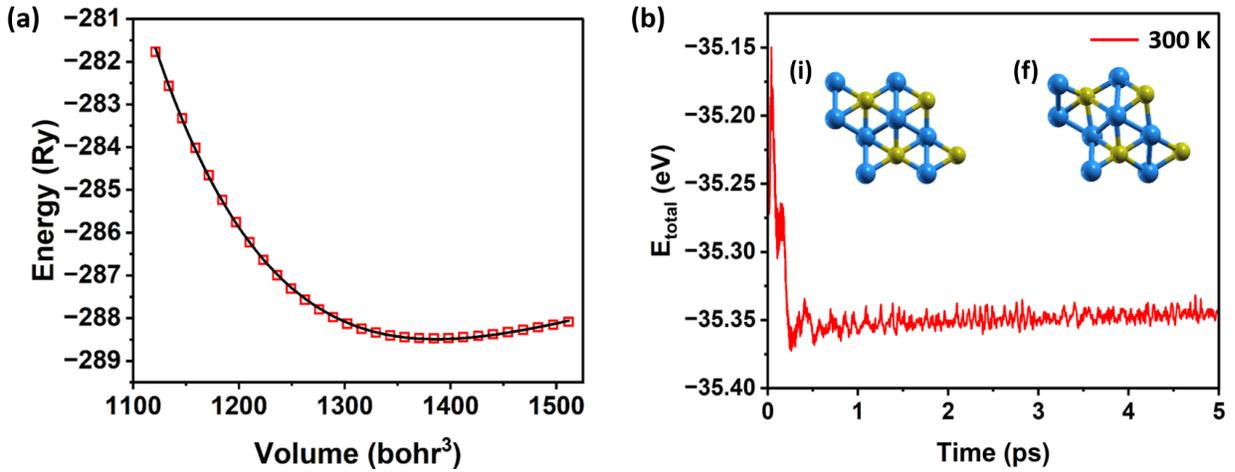

**Figure 2:** (a) Unit cell optimization of Be$_2$C monolayer. Here, the line represents the data fit to the Birch-Murnaghan equation of state [53]; (b) Fluctuation of total energy during AIMD simulations of the pristine Be$_2$C monolayer at 300K for 5 ps along with the top view. Insets (i) and (f) represent the initial and final geometries of the Be$_2$C monolayer, respectively.

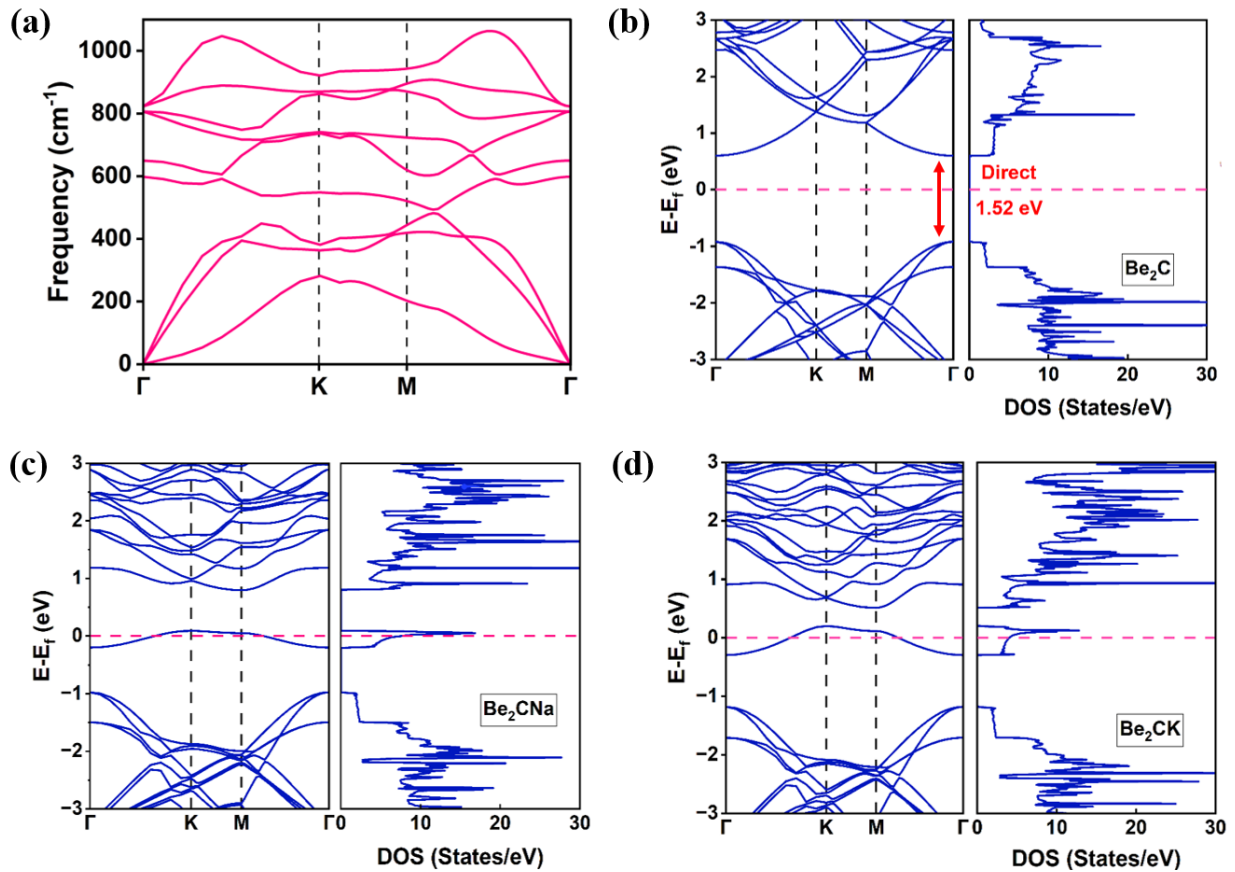

**Figure 3:** (a) Phonon dispersion curve for Be$_2$C monolayer. Electronic band structures and density of state profiles for (b) pure (c) Na-adsorbed (d) K-adsorbed Be$_2$C monolayer.

## Alkali Metal Atom Adsorption

To investigate the adsorption of alkali metal atoms (Na and K) on a Be$_2$C monolayer, first, we identify the possible adsorption sites, as marked in Fig. 1(a). A total of six potential adsorption sites are taken into consideration: three atomic sites (1) bottom beryllium (site-A), (2) top beryllium (site-B), (3) carbon (site-C); three bridge positions (4) Be-Be bridge site (black cross), (5) bottom Be-C bridge site (red cross), and (6) top Be-C bridge site (green cross). We conduct extensive structural optimizations for every adsorption site. We have observed that when we try to relax the adatom at the bridge site, it changes its position to site-A or site-C to achieve stability. So, in the present study, we have discussed only three sites: A, B, and C. Comparing the adsorption energies ($E_{ad}$), as determined using Eq. (2), a quantitative analysis of the adsorption behaviour and the favourable adsorption sites can be understood [56-58].

$$E_{ad} = (E_{Be_2C-M_x} - E_{Be_2C} - xE_M)/x \qquad (2)$$

Here, $E_{Be_2C-M_x}$ ($M$ = Na, K) and $E_{Be_2C}$ are the total energies of Be$_2$C with and without metal atom adsorption, respectively; $x$ corresponds to the number of metal atoms in the adsorption configurations and $E_M$ is the energy of an adsorbed alkali atom calculated for the same unit cell parameters as the Be$_2$C monolayer. There is no adsorption when the $E_{ad}$ is positive, indicating a repulsion between interacting species. The adsorption site is favoured by the negative values of $E_{ad}$. As seen in Fig. 4, All the estimated values of $E_{ad}$ is found negative for all A, B, or C site adsorption on Be$_2$C. Site A has the lowest $E_{ad}$ for both systems, making it the most suitable adsorption site for a single adatom. Adsorption energy values at site A are -0.71 eV and -0.76 eV for Na and K atoms, respectively.

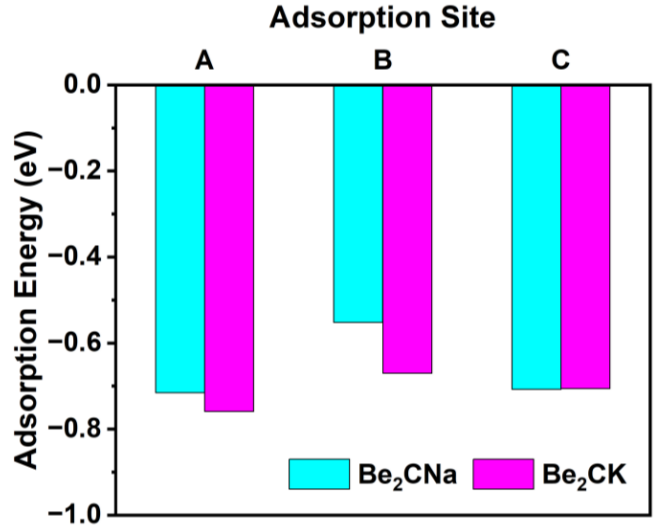

**Figure 4:** Adsorption energy for different sites of Be$_2$C monolayer with adsorbed Na and K atoms.

**Table 1:** Adsorption energy (in eV) and optimized adatom height, $h$ (in Å) for alkali-metal adsorption on the various sites of the Be$_2$C monolayer, as marked in Fig 1(a).

|  | Adsorption Energy (eV) | | | $h$ (Å) | | |
| --- | --- | --- | --- | --- | --- | --- |
|  | A | B | C | A | B | C |
| Be$_2$C – Na | -0.71 | -0.55 | -0.71 | 3.34 | 2.79 | 2.87 |
| Be$_2$C – K | -0.76 | -0.67 | -0.71 | 3.61 | 2.96 | 3.25 |

We have also optimized the height $h$ of the adsorbed alkali atom for different adsorption locations. These values of $h$ are summarized in Table 1 and are supported by the estimated values of $E_{ad}$. A high value of $h$ indicates a stronger adsorption.

**Diffusion of alkali metal ions on the 2D-Be$_2$C**

The mobility of the intercalating ions determines the charge-discharge rate, another essential characteristic that determines an electrode material's suitability for rechargeable batteries. Thus, we now focus on metal ion mobility and utilize the CI-NEB method on the Be$_2$C monolayer, as described in the Computational Details section. We have performed calculations for all three possible sites: A, B, and C.

Since the lowest value of $E_{ad}$ is obtained for site A among others [as mentioned in Fig. 1(a)], which makes it the most favourable, we consider three diffusion paths related to the symmetry of the 2D-Be$_2$C. These paths link two nearby A-sites, as shown in Fig. 5(a): (i) Path-I (A–C–A, red colour), (ii) Path-II (A–A, blue colour), and (iii) Path-III (A–B–A, green colour). Also, we have taken in to account site-C for this calculation, where metal ion migrates from one C site to another C site. Figure 5(d) displays possible paths: (i) Path-I (C–A–C, red colour), (ii) Path-II (C–C, blue colour), and (iii) Path-III (C–B–C, green colour). We could not achieve a minimum energy path between two B-sites (bottom Be) because site-B has a higher energy than the other two sites. So, metal atoms cannot pass through site A or site C.

Figures 5 (b) and (c) show the energy barrier profiles for site A along the different pathways corresponding to Na and K atoms, respectively. Path-1 exhibits the minimum diffusion barrier for K atoms, which goes from the top of the carbon atom, whereas, for Na atoms, the optimal pathway is path-3, passing through the site above the beryllium atom. The lowest diffusion barrier for site-A for metal atoms is 0.016 eV and 0.026 eV for Na and K metal atoms, respectively. Similarly, Figs. 5 (e) and (f) show the energy barrier profiles for site C. Since the adsorption energy of the Na atom is the same at sites A and C, the energy barrier profiles and energy barrier value are likewise the same (0.016 eV) in both cases. In the case of Be$_2$C-K, the K atom followed path-2 and path-3, but when we attempted to calculate path-1 (C-A-C), we could not achieve a minimum energy path because site-C has higher energy than site-A. The energy barrier profile for Be$_2$C-K, for site-C, is given in Fig. 5 (f), and the energy barrier for this case is 0.0108 eV. Figure 6 (a, b) shows the diffusion barrier values for three potential migration paths corresponding to Na and K atoms for sites-A and -C. Due to this small value of barrier energy, the adatoms can move quickly over the surface of 2D-Be$_2$C and easily participate in conduction, which can result in a better charge-discharge rate for these alkali MIBs.

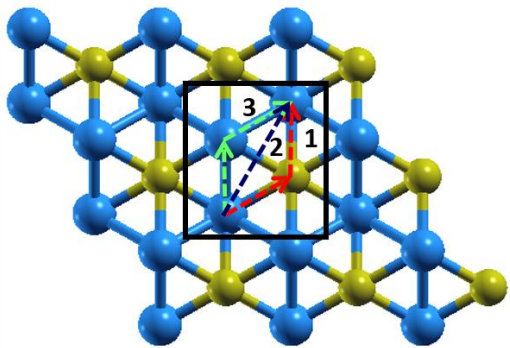
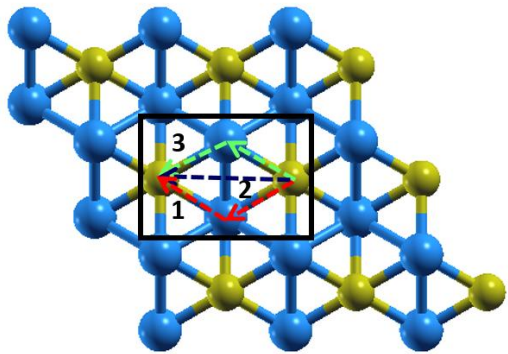
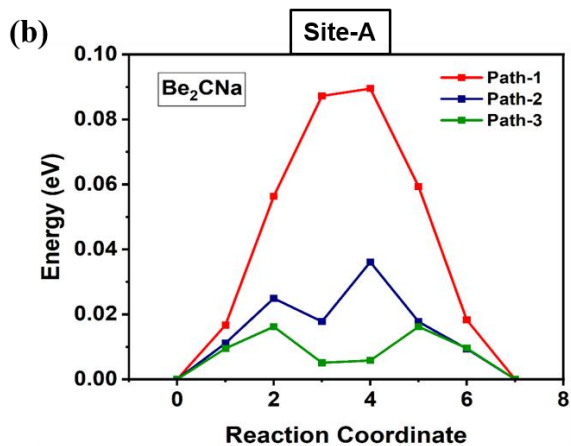
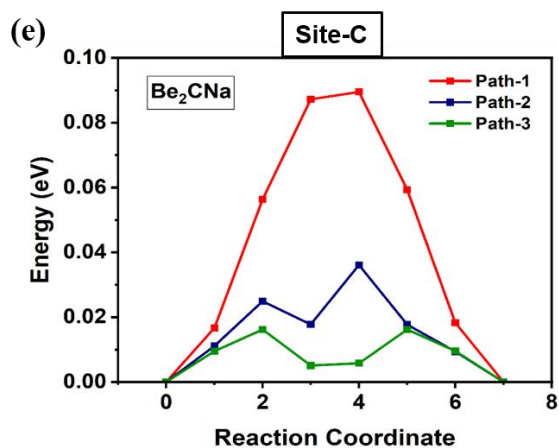
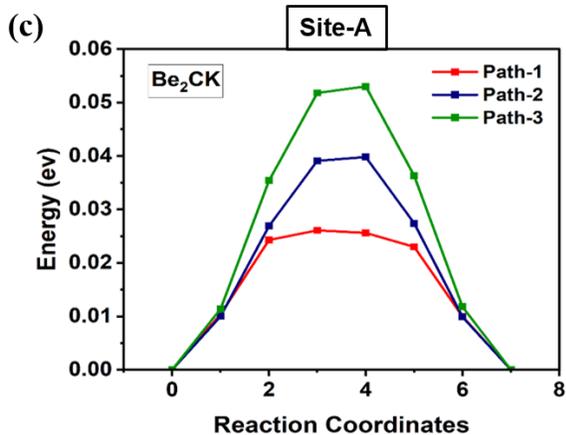
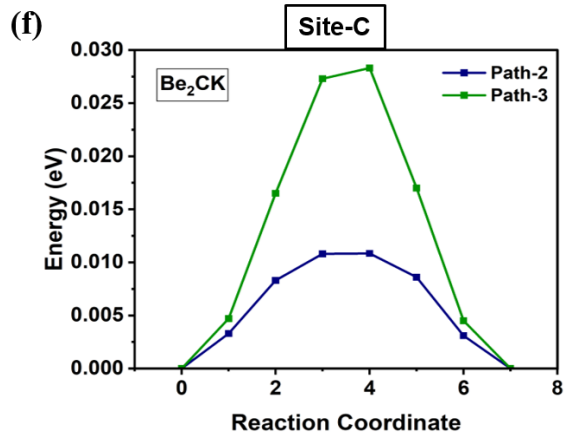

**Figure 5:** Schematic representation of three possible migration paths of alkali atom diffusion on the Be$_2$C monolayer (a) site-A and (d) site-C. (b, c) and (e, f) are the corresponding diffusion barrier profiles for Be$_2$C-Na and Be$_2$C-K, estimated for site-A and site-C, respectively.

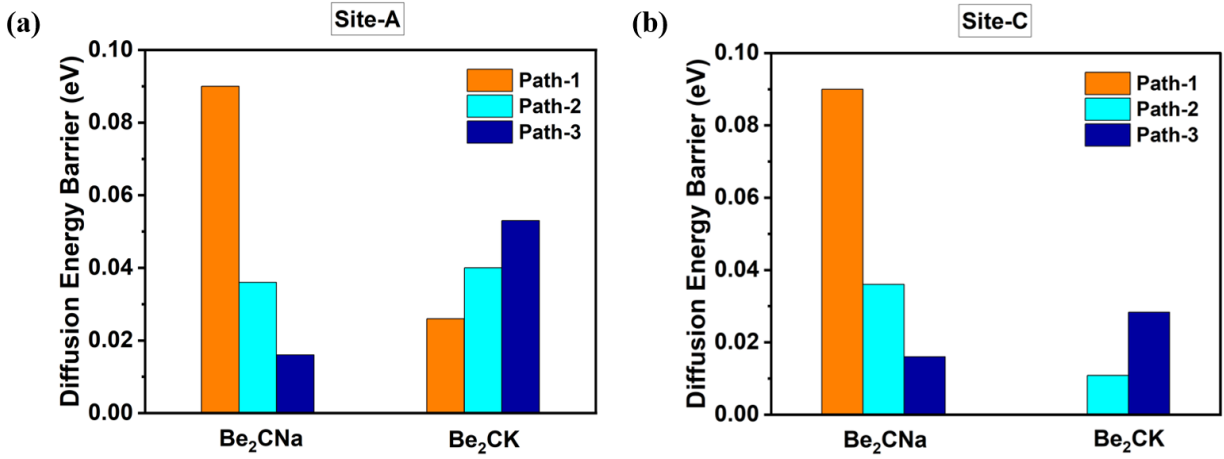

**Figure 6:** Diffusion energy barrier for three possible pathways for (a) site-A and (b) site-C for Be$_2$C-Na and Be$_2$C-K.

## Theoretical Open Circuit Voltage and Storage Capacity

The performance of electrode materials and batteries is generally understood by the open circuit voltage (OCV) and theoretical storage capacity (TSC). To introduce maximum adsorption of alkali atoms, we consider the adsorption on both sides of the 2D-Be$_2$C layer and then computed OCV and TSC for a $3 \times 3 \times 1$ supercell of Be$_2$C monolayer. The corresponding redox reaction for the Be$_2$C monolayer is given by Eq. (3).

$$Be_2C + xM^+ + xe^- \underset{Discharging}{\overset{Charging}{\rightleftharpoons}} Be_2CM_x \quad (3)$$

The OCV is computed using Eq. (4) for adsorbed Na and K atoms [52, 59, 60].

$$OCV = [-E_{Be_2CM_x} - E_{Be_2C} - xE_M^s]/xe \quad (4)$$

Here $E_M^s$ is the energy of a single isolated alkali atom and calculated as given by Eq. (5).

$$E_M^s = E_{surface-M + slab-M} - E_{slab-M} \quad (5)$$

In Eq. (5), $E_{slab-M}$ is the energy of the alkali metal slab, calculated for six layers of metal atoms. The first term on RHS is the system's energy containing a metal slab with one atom on the slab's surface. The vertical separation between the surface atom and slab is considered as $d$. By varying $d$, we have relaxed $d$ for both alkali atoms to find the optimized values of $E_M^s$. We observe that metal-slab and surface atoms have the lowest energy values at a $d$ value of 3.24 Å, and 4.05 Å for Na and K, respectively. The lowest $E_M^s$ is considered energetically the most stable. This value is used as the reference energy while computing OCV. We have calculated OCV using Eq. (4) for $x = 0.67$, 1 and 2. Figure 7(a-d) shows all considered configurations for different values of $x$. For $x = 0.67$, Na atoms exhibit a positional shift above site B and, site-C is found to achieve stability, and Be$_2$C-K is unstable for any of the three sites. For $x = 2$, stable adsorption sites-A and -C are possible for Na, whereas for K, only one stable site-C with 0.96 V of OCV. For $x = 1$, all sites are found stable for Na, but site-B is found to be unstable for K atom adsorption. In Table 2, computed

OCV is summarized for stable adsorption sites, and the cell, without any data, refers to an unstable site for that alkali metal. Materials with an OCV between 0.1 and 1.25 eV are categorized as negative electrode materials, whereas values over 1.25 eV are for the positive electrode [2, 61, 62]. Our results show that the OCV values for Na and K atoms are less than 1.25 V, indicating a promising candidature of the Be$_2$C monolayers for anode material in alkali MIBs.

TSC is related to ionic charges and provides information on how much electrical current is produced during the redox reaction of the battery. This is computed for the maximum adsorption of alkali atoms ($x$) per formula unit, as given by Eq. (6),

$$C_M = \frac{xzF}{M_{Be_2C}} \qquad (6)$$

where $M_{Be_2C}$ is the corresponding atomic mass, $F$ is the Faraday constant, and $z$ is the valence number, which is 1 in our case. From Eq. (6). A TSC value of around 1785 Ah/kg is estimated for $x = 2$. Here, $x = 2$ means that two metal atoms, above and below, are adsorbed in a unit cell.

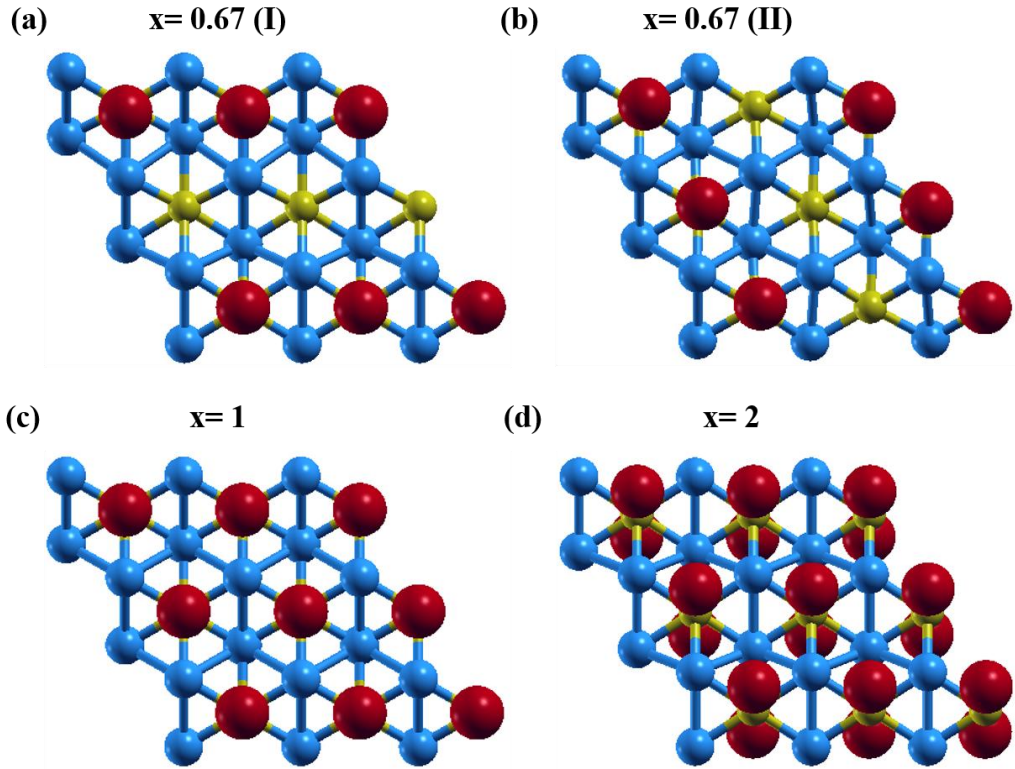

**Figure 7:** Structural representation of Be$_2$C-$M_x$ monolayers with different concentrations (a) $x = 0.67$ (I), (b) $x = 0.67$ (II), (c) $x = 1$, and (d) $x = 2$ (for site-C).

**Table 2** OCV values in voltage for Be₂C-Na and Be₂C-K monolayers for different values of $x = 2, 1, 0.67$. The cells without any data refer to an unstable situation.

| $x$ | Be$_2$C – Na | | | Be$_2$C – K | | |
|---|---|---|---|---|---|---|
| | A | B | C | A | B | C |
| 2 | 0.43 | - | 0.37 | - | - | 0.96 |
| 1 | 0.42 | 0.50 | 0.40 | 0.96 | - | 0.99 |
| **0.67 (I)** | 0.27 | - | 0.27 | - | - | - |
| **0.67(II)** | 0.27 | - | 0.27 | - | - | - |

## Conclusion

Based on DFT calculations, we have systematically examined the possible application of 2D-Be₂C as an anode material for metal-ion batteries. We have optimized the structure and phonon band structure with no negative frequency, indicating the stability of 2D-Be₂C. Semiconducting Be₂C monolayer with a direct bandgap of 1.52 eV changes to metallic after alkali metal adsorption, which can warrant better electron conduction. At the same time, adsorption of alkali-metal ions, a negligible change in the volume of the unit cell, is observed. Negative adsorption energy is computed for both alkali metal atoms, which secures the stability of the adsorbed atom/2D-Be₂C system. Our result shows that the value of the diffusion barrier is meagre: 0.016 eV (for Na) and 0.026 eV (for K), which is advantageous for the improved charging-discharging performance of alkali metal-ion batteries. Low open circuit voltage for our structures and high storage capacity suggests the potential usage of these 2D-Be₂C monolayers.

**Acknowledgement**
HP thanks the Science and Engineering Research Board (SERB), Govt. of India, for the research grant (ECR/2017/001612) and SVNIT Surat for the Institute seed Money grant 2021-22/DOP/04.

**Conflict of Interest**
The authors declare no conflict of interest.

**Data Availability Statement**
The data that support the findings of this study are available from the corresponding author upon reasonable request.